\newcommand{\beq}{\begin{equation}}
\newcommand{\eeq}{\end{equation}}
\newcommand{\beqa}{\begin{eqnarray}}
\newcommand{\eeqa}{\end{eqnarray}}
\newcommand{\gaa}{g_\parallel}
\newcommand{\gab}{g_\perp}
\newcommand{\Uaa}{U_\parallel}
\newcommand{\Uab}{U_\perp}

\newcommand{\ga}[1]{g_{#1\parallel}}
\newcommand{\gb}[1]{g_{#1\perp}}
\documentstyle[aps,prl,preprint,amsfonts,epsf]{revtex}
\begin{document}
\def\dfrac#1#2{{\displaystyle{#1\over#2}}}

%\twocolumn[\hsize\textwidth\columnwidth\hsize\csname @twocolumnfalse\endcsname

\title{
Superconducting Fluctuations in a Multi-Band 1D Hubbard Model
}

\author{
M.I.\ Salkola and A.V.\ Balatsky$^\ast$
}

\address{
 Theoretical Division, Los Alamos National Laboratory,
Los Alamos, New Mexico 87545
}

\date{March 1, 1994}
\maketitle

\begin{abstract}
A renormalization-group and bosonization approach for a multi-band Hubbard
Hamiltonian in one dimension is described. Based on the limit of many
bands, it is argued that this Hamiltonian with bare repulsive electron-electron
interactions is scaled under specific conditions to a model in which
superconducting fluctuations dominate.

\


\end{abstract}


%]

%\onecolumn
\newpage
%\narrowtext

The Hubbard model plays a paradigmatic role in understanding many-body
correlation effects. Stimulated by the discovery of high-$T_c$
superconductivity,
it has been subject of renewed interest and its multi-band versions with
purely repulsive interactions have been examined as possible realizations
of superconductivity \cite{KA}. However, the first study of multi-band
Hubbard models date back to Van Vleck \cite{Vleck} who evoked intra-atomic
interactions between orbitally degenerate states as an explanation for
ferromagnetism. In general, these models are fascinating because, due to
orbital degeneracy, they incorporate a variety of intriguing physical
phenomena, including, for example, the coexistence of orbital superlattice
and magnetic orderings \cite{more}. One potential application of multi-band
models could be doped fullerenes --- e.g., C$_{60}$ with three- and five-fold
degenerate $t_{1u}$ and $h_u$ bands --- where the consequences of band
degeneracy
are certainly important in understanding the physics of these novel materials.

While the above studies mostly consider narrow bands or only a few degenerate
bands within mean-field approximation, Muttalib and Emery \cite{Mutta}
studied recently a two-band model of spinless electrons in
one dimension by abelian bosonization and renormalization-group
analysis. Interestingly, they found that superconductivity can exist
with repulsive electron-electron interactions. They also considered
the question of decoupling of dynamical degrees of freedom which they found
to occur only at some specific values of coupling constants.

Motivated by the above examples, we focus in this paper on a novel, degenerate
multi-band Hubbard model for fermions
with spin degrees of freedom. We first derive one-loop renormalization-group
equations and
analyze their behaviour.  Solvable limits of this model are then studied
by bosonization. We argue that, for a particular choice of parameters, the
model exhibits enhanced superconducting fluctuations. We conclude by noting
that
the decoupling of dynamical degrees of freedom does not occur in general,
in contrast to the one-band Hubbard model.

Specifically, our starting point is the Hamiltonian
\beq
H=-t\sum_{n\alpha\sigma}(c^\dagger_{n+1\alpha\sigma}c_{n\alpha\sigma} + h.c.)
    + \mbox{$1\over2$} \sum_{n\alpha\beta} U_{\alpha\beta} \rho_{n\alpha}
 \rho_{n\beta},
\eeq
where $c_{n\alpha\sigma}$ is the fermion operator for an electron of spin
$\sigma$ with band index $\alpha$ and $\rho_{n\alpha} =
\sum_\sigma c^\dagger_{n\alpha\sigma}c_{n\alpha\sigma}$ at site $n$.
We associate the band degeneracy to colour degrees of freedom and spin
degeneracy to flavour degrees of freedom; the corresponding indices take
values $\alpha=1,\ldots,N_C$ and $\sigma=1,\ldots,N_F$. $U_{\alpha\beta}$
is the on-site electron-electron interaction which takes into account
possible couplings between various bands, and $t$ is the hopping amplitude
which has been assumed to be the same for all the bands. In the absence of
electron-electron interactions, the model has the U$(N_C\times N_F)$ symmetry
which is preserved for $U_{\alpha\beta}\equiv U$. By including more than one
band and allowing the broken colour symmetry,
interesting possibilities arise, one of which is enchanced
superconducting fluctuations, as seen below.  In the following,
we always assume the full spin-rotational symmetry.

Our primary results can be qualitatively explained by considering the
Hubbard interaction in the U($N_F$)-invariant form of
\beq
\mbox{$1\over2$}\Uab \sum_n \Big(\sum_\alpha \rho_{n\alpha}\Big)^2 +
 \mbox{$1\over2$}(\Uaa-\Uab) \sum_{n\alpha} \rho_{n\alpha} \rho_{n\alpha},
\eeq
where we have assumed that the diagonal interactions can be parametrized
by $U_\parallel$ and the off-diagonal interactions by $U_\perp$. Subsequently,
we will be mostly interested in cases where $U_\perp \ge 0$ while
$U_\parallel$ may have an arbitrary sign.
Our results are: ($i$) In the limit of $N_C\rightarrow \infty$, $U_\perp$
scales
to zero while $U_\parallel - U_\perp$ is invariant under the scaling.
Thus, the model will scale to $N_C$ independent systems whose
on-site electron-electron interactions are either attractive or
repulsive, depending on the sign of $U_\parallel - U_\perp$.
This is expected, because the mean-field approach should work in this
limit and so the first term in Eq.~(2) is qualitatively unimportant.
($ii$) For finite but large $N_C$, the system again scales so that $U_\perp$
decreases and renormalized diagonal interaction may be positive or
negative. However, if initially $U_\parallel < U_\perp$, there exists a
limiting value for the effective $U_\perp$ after which further scaling causes
it to increase. This critical value scales as $1/N_C$.

{\it Current Algebra and the Action.}
We will use the current-algebra formalism in the calculations
because it provides a compact notation and allows operator-product
expansions to be used efficiently in deriving the scaling equations as
short-distance degrees of freedom are integrated out. Since
we are interested in low-energy and long-wavelength phenomena, the spectrum is
linearized at the Fermi energy and the fermionic degrees of freedom are
expressed
by slowly varying fields $\psi_{\alpha\sigma}(x_\pm)$, where the left ($+$)
and right ($-$) moving electrons are labeled according to their arguments,
$x_\pm = t \pm x$. We define $J^{\sigma\sigma'}_{\alpha\alpha'}(x_\pm)=
\hbox{:$\psi^\dagger_{\alpha\sigma}(x_\pm) \psi_{\alpha'\sigma'}(x_\pm)$:}$
which present the left and right moving currents; the upper (lower) indices
of the current refer to the flavour (colour) degrees of freedom. The colons
denote
normal orderings with respect to the filled Fermi sea of the noninteracting
system. Flavour and colour currents are
then defined naturally as
$J^a(x_\pm)= \tau^a_{\sigma\sigma'} J^{\sigma\sigma'}_{\alpha\alpha}(x_\pm)$
and
$J^A(x_\pm)= T^A_{\alpha\alpha'} J^{\sigma\sigma}_{\alpha\alpha'}(x_\pm)$,
respectively (hereafter a sum over
repeated indices is implied). Hermitian matrices $\tau^a$
and $T^A$ form a Lie algebra of SU($N_F$) and SU($N_C$) groups.
Note that it is convenient to add the identity matrix to the above sets of
generators; for example $\tau^0 = {\bf 1}/\sqrt{N_F}$ \cite{Lie}.
Therefore, both currents actually give the U(1) current with $a=A=0$,
whereas $J^a$ and $J^A$ are the true SU($N_F$) and SU($N_C$) currents
for nonzero $a$ and $A$. It is easy to compute the current commutation
relations, % \cite{commut}
equivalent to the Kac-Moody algebra.
However, more useful is the corresponding operator-product expansion,
\beqa
&J^{\sigma\sigma'}_{\alpha\alpha'}&(z_\pm)J^{\mu'\mu}_{\beta'\beta}(w_\pm)
=
\mp
{\delta_{\alpha\beta}\delta_{\alpha'\beta'}\delta_{\sigma\mu}\delta_{\sigma'\mu'}
\over 4\pi^2 (z_\pm-w_\pm)^2 }  \label{eq:ope} \\
& & \quad\quad\quad\quad \quad\quad\quad\quad \mbox{ } +
{J^{\sigma\mu}_{\alpha\beta}(z_\pm)\delta_{\alpha'\beta'}\delta_{\sigma'\mu'}-
J^{\mu'\sigma'}_{\beta'\alpha'}(z_\pm)\delta_{\alpha\beta}\delta_{\sigma\mu}
\over 2\pi ( z_\pm - w_\pm) } + \cdots. \nonumber
\eeqa
This follows from the commutator and can be shown by means of
the conformal symmetry. Here we assume radial quantization where the
spatial dimension is compactified by mapping the space and time coordinates
to a cylinder of radius $2\pi$; $z_\pm = \exp(t \pm ix)$.
Below, we continue to use the flat space-time coordinates for the action
but do the actual calculations in the complex plane.

The free part of the action is \cite{ia}
\beq
S_o= {\pi \over N_C+N_F} \sum_{s=\pm} \int d^2x  [J^a(x_s)J^a(x_s) +
J^A(x_s)J^A(x_s)],
\eeq
and the action describing the electron-electron interaction is
\beqa
S_I&=& \int d^2x \left(-U^{(1)}_{\alpha\beta}
J^{\sigma\mu}_{\alpha\beta}(x_+)J^{\mu\sigma}_{\beta\alpha}(x_-)
+  U^{(2)}_{\alpha\beta}
J^{\sigma\sigma}_{\alpha\alpha}(x_+)J^{\mu\mu}_{\beta\beta}(x_-)
\right. \nonumber\\
& & + \mbox{$1\over2$}U^{(3)}_{\alpha\beta} [
\hbox{:$\psi^\dagger_{\alpha\sigma}(x_+) \psi_{\alpha\sigma}(x_-)$:}\,
\hbox{:$\psi^\dagger_{\beta\mu}(x_+)\psi_{\beta\mu}(x_-)$:} + h.c.]
\label{eq:int}
\\
& & \left. +\mbox{$1\over2$}U^{(4)}_{\alpha\beta}
[J^{\sigma\sigma}_{\alpha\alpha}(x_+)J^{\mu\mu}_{\beta\beta}(x_+)
+J^{\sigma\sigma}_{\alpha\alpha}(x_-)J^{\mu\mu}_{\beta\beta}(x_-)]\right),
\nonumber
\eeqa
where $d^2x=dxdt$ and the units are defined so that the Fermi velocity becomes
$\hbar v_F=1$. In contrast to the Hubbard model,
we will allow the coupling constants $U^{(k)}$ to be independent, as implied
by various physically important interactions ---
for example, by the nearest-neighbour electron-electron interaction.

{\it Renormalization-Group Equations and the Flow.}
We follow the usual procedure in which the partition function, $Z=\int {\cal D}
[\psi] e^{-S}$, is regularized by introducing a ultraviolet cutoff $\ell$ for
short distances. As the short-distance degrees of freedom are integrated out
continuously by using the operator-product expansion (\ref{eq:ope}), the
coupling constants are correspondingly renormalized so that $Z$ remains
unchanged
\cite{ia,poly}. This procedure then gives the one-loop scaling equations for
the coupling constants:
%\begin{mathletters}
\beq
\begin{array}{llll}
\partial_\ell \ga{1}&= -(\gb{1}^2+\gb{3}^2)
&+&[(\gb{1}^2+\gb{3}^2) - \ga{1}^2]/N_C\\
\partial_\ell \gb{1}&= -(\gb{1}^2+\gb{3}^2)
 &+& [(\ga{2}- 2\ga{1} -\gb{2}+ 2\gb{1})\gb{1}
 + (2\gb{3}-\ga{3})\gb{3}]/N_C\\
\partial_\ell \ga{2}&= \hbox to 2truecm{ }& &
(\ga{3}^2 - \ga{1}^2)/2N_C\\
\partial_\ell \gb{2}&=  \hbox to 2truecm{ } & &
(\gb{3}^2 - \gb{1}^2)/2N_C\\
\partial_\ell \ga{3}&= -2\gb{1}\gb{3} &+&
[(2\ga{2}-\ga{1})\ga{3} + 2 \gb{1}\gb{3}]/N_C \\
\partial_\ell \gb{3}&= -2\gb{1}\gb{3}
&+& [(\ga{2}+\gb{2} - 2\ga{1} + 4\gb{1})\gb{3} -
\ga{3}\gb{1}]/N_C,
\end{array} \label{eq:b1}
\eeq
%\end{mathletters}
where $g_{k\parallel,\perp} = N_C U^{(k)}_{\parallel,\perp}/\pi$
($k=1,\ldots,4$) and $N_F=2$ is chosen. The coupling constants are functions of
the ultraviolet cutoff $\ell$; $\partial_\ell \equiv \partial/\partial\log
\ell$.
As usual, $g_{1\alpha}$ is the backward-scattering constant, $g_{2\alpha}$
is the forward-scattering constant, and $g_{3\alpha}$ is the Umklapp-scattering
constant ($\alpha= \, \parallel,\perp$). Because $g_{3\alpha}$ corresponds to
the
scattering processes which violate momentum conservation by
$4k_F$, it is important only if $4k_F$ is equal to the reciprocal lattice
constant;  $k_F$ is the Fermi wavevector which has been assumed to be
the same for all the bands.
We have ignored $g_{4\alpha}$ because its only effect is to
renormalize Fermi velocities; although this could be taken into account,
$g_{4\alpha}$ does not enter at one-loop order since $\oint dz/z^2 =0$.

Because the parameter space of the system is very large, we have investigated
the scaling behaviour of the system in two special cases:

(1) First,
we consider the system with the broken orbital $U(N_C)$ symmetry so that the
action is only $U(N_F)$ invariant. We further set the forward-scattering
coupling
temporally to zero, $g_{2\parallel,\perp} =0$, because the backward- and
Umklapp-scattering coupling constants are the most relevant ones in terms of
that their scaling behaviour is nontrivial. By letting initially
$g_{\parallel,\perp}=g_{k\parallel,\perp}$ ($k=1,3$), and
$g_{2\parallel,\perp}=0$, these properties are conserved by the
renormalization-group flow. Thus,
%\begin{mathletters}
\beq
\begin{array}{lll}
\partial_\ell \gaa &=& -2 \gab^2 + (2\gab^2-\gaa^2)/N_C, \\
\partial_\ell \gab &=& -2 \gab^2 + (4\gab-3\gaa)\gab/N_C;
\end{array} \label{eq:b2}
\eeq
%\end{mathletters}
with $N_F=2$.
It is easy to see from these equations that the renormalization-group flow may
change the sign of $\gaa$ but not the sign of $\gab$. In fact,
$\gab=0$ forms a fixed point for this coupling constant. This, and the line
$\gab=\gaa$ act as separatices:
($i$) First, let $\gab>\gaa \ge 0$. Now there is also another
relevant solution for the equation $\partial_\ell \gab=0$:
$\gab= - 3\gaa/(2N_C-4)$. This solution yields a turning point for the
scaling trajectories at which the initially decreasing
$\gab$ starts to increase. For large $N_C$, the turning value of $\gab$ scales
as $1/N_C$. Note that $\gaa$ decreases monotonically and, at the turning
point, $\gaa < 0$. ($ii$) In the region $\gaa\ge\gab\ge 0$, the system scales
towards a noninteracting model ($g_{\parallel,\perp}=0$), corresponding to the
$N_C$ free fermion theories. ($iii$) Finally, $\gab<0$ leads always to the
strongly attractive interactions: $\gaa$ and $\gab$ decrease without bound
at one-loop order.

To illustrate the general behaviour of Eqs.~(\ref{eq:b2}), we integrate
them numerically and plot the renormalization-group flow for $N_C=10$; see
Fig.~1. For $\gab>\gaa >0$, it clearly shows the tendency
of the renormalization-group flow to lead to an effective Hamiltonian which
contains weakly coupled one-band Hamiltonians with intraband
electron-electron attractions. It is then convenient to diagonalize
each of these Hamiltonians separately and to take into account the
off-diagonal interactions perturbatively. However, this assumption relies on
the
fact that the ``unperturbed'' system has a gap the perturbation expansion to
work. These results apply to Eqs.~(\ref{eq:b1}) whose qualitative behaviour
is similar for finite $g_{2\parallel,\perp}$.

(2) We next consider the model with the full
U$(N_C\times2)$ symmetry: $U^{(k)}_{\alpha\beta}=U^{(k)}$, $k=1,\ldots,4$.
In this case, the interaction term becomes
\beqa
S_I= \int&d^2x& \Big( -U^{(1)} J^{aA}(x_+)J^{aA}(x_-) +
 U^{(2)} J(x_+)J(x_-)   \nonumber \\
  \mbox{ }&+&\mbox{$1\over2$} U^{(4)} [J(x_+)J(x_+)
+ J(x_-)J(x_-)] \label{eq:int2}\\
      &+& {\rm Umklapp} \Big), \nonumber
\eeqa
where  $J^{aA}(x_\pm) = \tau^a_{\sigma\sigma'} T^A_{\alpha\alpha'}
J^{\sigma\sigma'}_{\alpha\alpha'}(x_\pm)$
is the colour-flavour current and $J\equiv J^{00}$ is the U(1) current.
As an example, we assume that
the bare interactions are repulsive ($g_k>0$). In this case and at half
filling,
$g_1$ can scale to large negative values, $g_2$ is renormalized only weakly,
and
$g_3$ decreases and starts to increase approximately when $g_1$ becomes
negative. Note that the renormalized value of $g_3$ is always positive. It is
very difficult to draw any conclusions, even at $g_1=0$ for
$g_3\ne0$ (the half-filled case), although the interesting feature is that the
backward-scattering
coupling constant $g_1$ scales towards large negative values which may suggest
enhanced density-wave fluctuations in colour or flavour channel.
In general, an analytical solution does not seem to be possible when
$g_1$ and $g_3$ are the most dominant coupling constants.

{\it Bosonization.} We now turn to a limit where the backward and Umklapp
terms are either zero or scale to zero, so that the multi-band model is
exactly solvable by bosonization \cite{nona,Eme}. In particular, we consider
the U$(N_F)$-invariant model where the U$(N_C$) symmetry is broken by
letting $U_\parallel\ne U_\perp$.
For our purposes, abelian bosonization \cite{Eme} is sufficient.
At zero temperature, the most dominant instabilities are either $2k_F$
density-wave (DW) or superconducting (SC) instabilities whose
correlation functions \cite{mu} have power-law singularities at low energies
and small momenta with exponents
\begin{mathletters}
\beqa
\mu_{\rm DW}&=& 2(1-\delta\theta)/N_F + 2(\delta\theta-\bar{\theta})/N_CN_F,\\
\mu_{\rm SC}&=&2(1-1/\delta\theta)/N_F + 2(1/\delta\theta
-1/\bar{\theta})/N_CN_F,
\eeqa
\end{mathletters}
where $\delta\theta^2= (1+\delta u_4-\delta u_2)/(1+\delta u_4+\delta u_2)$
and $\bar{\theta}^2= (1+\bar{u}_4-\bar{u}_2)/(1+\bar{u}_4+\bar{u}_2)$. We have
defined $\delta u_k=N_F (U^{(k)} _{\parallel}-U^{(k)}_{\perp})/4\pi$
and $\bar{u}_k= N_F [U^{(k)}_{\parallel} +(N_C-1)U^{(k)}_{\perp}]/4\pi$,
($k=2,4$).
In these expressions, $U^{(k)}_{\parallel,\perp}$ are the renormalized
values of the forward-scattering constants. Clearly density-wave
fluctuations always dominate over superconducting fluctuations for
$U^{(k)}_{\parallel}=U^{(k)}_{\perp}>0$ and vice versa. However, if
$0\le U^{(2)}_{\parallel} <  U^{(2)}_{\perp}$, superconducting
fluctuations win,
provided that $N_C>(\delta\theta/\bar{\theta} -1)/(\delta\theta -1)$ and
$U^{(2)}_{\perp}\le U^{(4)}_{\perp}$, where the latter condition is needed to
guarantee a sensible large-$N_C$ limit.

In contrast, if the large-$N_C$ limit
is taken so that
$g_{k\parallel,\perp}\equiv N_CN_FU^{(k)}_{\parallel,\perp}/2\pi$
are kept constant, the condition, $0\le g_{2\parallel}< g_{2\perp}$, is no
longer
sufficient to have diverging superconducting correlation functions at low
energies
and small momenta. As an example, let
$g_{\parallel,\perp}=g_{k\parallel,\perp}$ ($k=2,4$) and consider the
weak-coupling
limit: $g_{\parallel,\perp}\ll 1$. We find that the superconducting correlation
functions are the most singular ones for the repulsive interactions and $N_C
\gg 1$,
if the condition $g_\parallel < g^2_\perp/2 + {\cal O}(g^3_\perp)$ is satisfied
\cite{compl}.

 It is interesting to note the connection between the above consideration and
the renormalization-group approach in the case of the repulsive interactions:
they
both imply that the necessary condition for superconducting fluctuations
to dominate is $U_{\parallel}<U_{\perp}$ and $N_C$ to be large enough.

{\it Colour-Flavour Separation.} The scaling equations show no apparent
separation
into independent sectors. This not so surprising if we consider the model with
the
\hbox{U$(N_C\times N_F)$} symmetry.
This coupling of the colour and flavour
degrees of freedom --- a manifestation of the fact that the colour and flavour
currents are not independent dynamical variables --- follows because the
colour-flavour current commutes neither with the colour nor with the flavour
current \cite{com}. However, away from half filling, the U(1) current describes
an independent dynamical degree of freedom; $J$ commutes with the other
currents.
The coefficient multipling the $JJ$ term in Eq.~(\ref{eq:int2}) is
$U^{(c)}= U^{(2)} - U^{(1)}/N_CN_F$. Consequently, in the one-loop
renormalization-group equations (\ref{eq:b1}),
$g_c\equiv N_CN_F U^{(c)}/2\pi$ is indeed decoupled from the other degrees of
freedom. Moreover, it is a scaling invariant. In the one-band Hubbard
model, the spin-charge separation is evident even at half-filling, because
the Umklapp term is a chiral SU($N_F$) singlet \cite{Affleck}.

{\it Relation to the Muttalib-Emery Approach \cite{Mutta}}.
Our approach differs in two
respects from that of Ref.~\cite{Mutta}: we consider an arbitrary number of
bands with the
spin degrees of freedom but ignore the interband Umklapp scattering. The
interband
Umklapp scattering would allow versatile possibilities; for example, one could
include
terms such as $U^{(5)}_{\alpha\beta} \int \! d^2x \,
J^{\sigma\sigma}_{\alpha\beta}(x_+)J^{\mu\mu}_{\alpha\beta}(x_-)$.
Since the scaling equations have to be invariant under the phase transformation
$\psi_{\alpha\sigma}\rightarrow e^{i\theta_\alpha} \psi_{\alpha\sigma}$,
$U^{(5)}_{\alpha\beta}=0$ is a fixed point and the other scaling equations
contain only even powers of $U^{(5)}_{\alpha\beta}$. The question remains
whether these coupling constants are (marginally) relevant. Because they do not
originally appear in the Hubbard model, we have assumed that their bare values
are negligible and that they do not change the scaling behaviour qualitatively.
On the other hand, in the model considered in Ref.~\cite{Mutta}, they were
found
to be important. Our motivation was to illustrate that even the Hubbard model
can show enhanced superconducting fluctuations if the spin and multiple bands
are assumed.

{\it Conclusion.} We have shown that the existence of multiple bands can lead
to
enhanced superconducting fluctuations. Based on the one-loop
renormalization-group
analysis, we found that the renormalized diagonal electron-electron
interactions
are attractive and  the off-diagonal electron-electron interactions are of
order
${\cal O}(N_C^{-1})$, if  bare values of the coupling constants are such that
$U_\parallel < U_\perp$.
This is consistent with the result obtained by studying the Hamiltonian in
the exactly solvable limit.

A.V.B.~acknowledges the support of J.R.\ Oppenheimer fellowship.
This work was supported by the U.S.\ Department of Energy.

%\newpage

\figure{\noindent FIG.1 \ Scaling trajectories in the ($\gaa, \gab$)-plane
for the half-filled, $N_C$-band Hamiltonian for $N_C=10$, as
given by Eq.~(\ref{eq:b2}) ($N_F=2$). The equation $\gab= - 3\gaa/(2N_C-4)$
is shown by a dashed line.}


\begin{thebibliography}{99}
\bibitem{Lie} The SU($N_F$) generators are renormalized so that
${\rm Tr}\,\tau^a\tau^b = \delta^{ab}$ ($a,b=0,\ldots,N_F^2-1$). Other useful
relations are
$[\tau^a,\tau^b]=if^{abc}\tau^c$, and $\tau^a_{\sigma\sigma'}\tau^a_{\mu'\mu}
=\delta_{\sigma\mu}\delta_{\sigma'\mu'}$. Note that $\tau^0= {\bf
1}/\sqrt{N_F}$.
Similar equations hold for the SU($N_C$) generators $T^A$
($A=0,\ldots,N_C^2-1$), where the structure factors are now denoted by
$f^{ABC}$.
\bibitem{mu} These correlation functions, which are the most singular
for the present U$(N_F)$-invariant model, are described either by the
operators $A^{a}_{\rm FSC}(x,t)= \psi_{\alpha\sigma}(x_+) \tau^a_{\sigma\mu}
\psi_{\alpha\mu}(x_-)$ (flavour-superconducting) or by $A_{\rm SDW}(x,t)=
\psi^\dagger_{\alpha\sigma}(x_+) \psi_{\alpha\sigma}(x_-)$
(scalar-density-wave)
and $A^{a}_{\rm FDW}(x,t)= \psi^\dagger_{\alpha\sigma}(x_+)  \tau^a_{\sigma\mu}
\psi_{\alpha\mu}(x_-)$ (flavour-density-wave) ($a\ne 0$). For the
U$(N_C\times N_F)$-invariant model, we have also other, equally divergent
correlations functions at low energies, as determined by the U$(N_C\times N_F)$
generators. (If $N_CN_F\ne 2$, the operator for scalar-superconducting
fluctuations is more involved and it is not considered here.) At zero
momenta (relative to $2k_F$), the low-energy behaviour of the correlation
functions is described by the form of $\omega^{-\mu}$ which also defines
the corresponding exponent $\mu$.
\bibitem{compl} If the weak-coupling limit is relaxed, this condition
will become $g_\parallel < g_\perp +2 -2\sqrt{1+g_\perp}$. Note that
$\mu_{\rm DW,SC} = {\cal O}(N_C^{-1})$, for $N_C\gg 1$ and fixed $g$'s.
\bibitem{com} For example, $[J^{a}(x_\pm),J^{bB}(y_\pm)]= if^{abc}J^{cB}(x_\pm)
\delta(x_\pm-y_\pm) \pm {1 \over 2\pi i} \delta^{ab}\delta^{0B}\partial_x
\delta(x_\pm - y_\pm)$.
\end{thebibliography}
\end{document}